\documentclass[twocolumn,showpacs,preprintnumbers,amsmath,amssymb]{revtex4}
\newcommand{\be}{\begin{equation}}
\newcommand{\ee}{\end{equation}}

\newcommand{\ber}{\begin{eqnarray}}
\newcommand{\eer}{\end{eqnarray}}

\newcommand{\bra}{\langle}
\newcommand{\ket}{\rangle}

\newcommand{\bs}[1]{\ensuremath{\boldsymbol{#1}}}
\begin{document}

\title{Density functional theory for self-bound systems}
\author{Nir Barnea}
\email{nir@phys.huji.ac.il}
\affiliation{The Racah Institute of Physics, The Hebrew University,
91904 Jerusalem, Israel.\\
     Institute for Nuclear Theory, University of Washington, 98195 Seattle,
  Washington, USA}
\date{\today}

\begin{abstract}
The density functional theory is extended to account for self-bound systems.
To this end the Hohenberg-Kohn theorem is formulated for the intrinsic density 
and a Kohn-Sham like procedure for an $N$--body system is derived using the
adiabatic approximation to account for the center of mass motion.
\end{abstract}

\pacs{21.60.Jz,31.15.E-}
\maketitle

{\it Introduction --}
The formulation of a density functional theory (DFT) for self-bound systems is
a question of current interest in nuclear physics \cite{Engel06,Giruad06}. 
Nuclear application of the DFT through the Skyrme model or using
different density functionals is the only
viable way to study heavy nuclei. 
The basic physical quantity
in the DFT is the single particle density. This quantity, being well defined
for an $N$--body system being localized by an external field is essentially
zero for a 
freely moving self-bound system \cite{Engel06}. Therefore, when treating
finite self-bound many-particle systems the DFT should be modified to account
for center of 
mass motion. This modification can be looked for along two possible lines,
either to add an 
external potential, confine the system so that the DFT applies and 
remove the center of mass 
effects at the end of the calculation \cite{Giruad06}. The other possibility
is to derive the Hohenberg-Kohn (HK) theorem \cite{HK} for the intrinsic density
and construct a Kohn-Sham \cite{KS} like procedure for calculating the density
and 
energy \cite{Engel06}. 
In this manuscript we shall follow the second path. We shall first derive the
HK theorem for the intrinsic density. This theorem, as the original HK theorem,
is a specific case of a more general theorem by Valiev and Fernando who proved
the existence of an energy functional 
for any Hermitian operator
\cite{Valiev97,Engel06}. Here we rederive this proof using our notation for
completeness and 
clarity. Once we have the HK theorem at our disposal, we follow the KS
procedure, and   
construct a KS-like Schr\"odinger equation, which can be separated into 
a single particle equation using a Born-Oppenheimer type
approximation for the ``slow'' center of mass coordinate. In this last step
one is forced to introduce some approximations since the desired KS orbitals
break the translation invariance.

For clarity we shall restrict our discussion to a particle system interacting
via two--body forces which only depend on the relative distance between the
particles. Furthermore, in deriving the KS like equations we shall follow the
Hartree rather than the Hartree-Fock scheme.

{\it The HK theorem for the intrinsic density --}
Consider a system of $N$ particles moving under the influence of a two--body
potential 
\begin{equation}
 U = \sum_{i\neq j}^{N} u(\bs{r}_i-\bs{r}_j)
\end{equation}
and an intrinsic ``one''--body potential
\begin{equation}
 V_{in} = \sum_{i}^{N} v_{in}(\bs{r}_i-\bs{R})
\end{equation}
where $\bs{R} = \frac{1}{N}\sum_i^N \bs{r}_i$ is the center of mass coordinate.
The Hamiltonian describing the internal motion of the system is 
\begin{equation}
 H = T_{in} + U + V_{in}
\;,
\end{equation}
where the intrinsic kinetic energy operator is given by
\begin{equation}
 T_{in} = \sum_{i}^N \frac{\bs{p}_i^2}{2} - \frac{\bs{P}^2}{2 N} \;.
\end{equation}
Here $\bs{P} = \sum_{i}^N \bs{p}_i$ is the center of mass momentum.
Following HK we shall assume a nondegenerate ground state $\Psi$. The intrinsic
translational invariant density 
\begin{equation}
 n_{in}(\bs{r})=\bra \Psi | \sum_{i}^N \delta(\bs{r}_i-\bs{R}-\bs{r}) |\Psi\ket
\end{equation}
is clearly a functional of $v_{in}(\bs{r})$. 
Following the HK method we can prove
that conversely (up to a constant) $v_{in}(\bs{r})$ is a functional of
$n_{in}(\bs{r})$. 
We start by assuming that another potential $v'_{in}(\bs{r})$ with a ground
state $\Psi'$  
gives rise to the same intrinsic density $n_{in}(\bs{r})$. The two wave
functions 
$\Psi,\Psi'$ must be different unless $v_{in}-v'_{in}=const$, since they
correspond to two 
different Hamiltonians $H$ and $H'=T_{in}+U+V_{in}'$. Utilizing the
variational principle one gets
\begin{equation}
  E'=\bra \Psi' |H'|\Psi'\ket < \bra \Psi| H' | \Psi \ket
    = \bra \Psi| H + V_{in}'-V_{in} | \Psi \ket
\end{equation}
or
\begin{equation}
  E' < E + \int d\bs{r}\, n_{in}(\bs{r})
                \left(v'_{in}(\bs{r})-v_{in}(\bs{r})\right)\;.
\end{equation}
Interchanging primed and unprimed quantities, we get
\begin{equation}
  E < E' + \int d\bs{r}\, n_{in}(\bs{r})
                \left(v_{in}(\bs{r})-v'_{in}(\bs{r})\right)\;.
\end{equation}
Adding the last two equations we reproduce the famous inconsistency
\begin{equation}
  E+E' < E'+E \;.
\end{equation}
Thus we have shown that $v_{in}(\bs{r})$ is a functional of the intrinsic
density $n_{in}(\bs{r})$. Since, $v_{in}(\bs{r})$ fixes $H$ we see that the
intrinsic $N$--body ground state is a functional of
$n_{in}(\bs{r})$. 

Since $\Psi$ is a functional of the intrinsic density we can define the
functional
\begin{equation}
 F[n_{in}(\bs{r})] = \bra \Psi | T_{in}+ U | \Psi \ket \;,
\end{equation}
which is a universal functional valid for any number of particles as we can
repeat the proof for any number of particles $N$ which by itself is a
functional of 
$n_{in}(\bs{r})$. The HK variational principle for the ground state energy
\begin{equation}
  E[n_{in}(\bs{r})]\equiv \int d\bs{r}\, n_{in}(\bs{r}) v_{in}(\bs{r}) 
      + F[n_{in}(\bs{r})] 
\end{equation}
follows trivially from the original proof \cite{HK}.

{\it The KS equations --}
Once it has been established that the energy of the finite self--bound
many--body system is a functional of the intrinsic density the next step is
to construct extended KS equations for such system. 
The ground state energy of the $N$--body system interacting via two--body
potential $u$ can be written in the form
\begin{equation}\label{E_gs}
  E = \frac{1}{2}\int \int d\bs{r}\, d\bs{r}' n_{in}(\bs{r})
      u(\bs{r}-\bs{r}')n_{in}(\bs{r}')+ G[n_{in}(\bs{r})]\;.
\end{equation}
where $G$ is a universal functional of the intrinsic density. Following KS we 
write
\begin{equation}
 G[n_{in}(\bs{r})] = T_s[n_{in}] + E_{xc}[n_{in}]
\end{equation}
where $T_s[n_{in}]$ is the intrinsic kinetic energy of a system of particles
interacting with a ``one''--body potential $v_{in}[n_{in}]$, and
$E_{xc}[n_{in}]$ is the exchange-correlation functional.
Due to the HK
variational principle for the energy, Eq. (\ref{E_gs}), we obtain the equation
\begin{equation}\label{varE}
   \int d\bs{r}\, \delta n_{in}(\bs{r})\left[
       \frac{\delta T_s[n_{in}] }{\delta n_{in}(\bs{r})}
     + \varphi(\bs{r})
    \right]=0 \;,
\end{equation}
subject to the condition
\begin{equation}\label{varN}
  \int d\bs{r}\, \delta n_{in}(\bs{r})=0 \;.
\end{equation}
Here
\begin{equation}\label{1b_poten}
  \varphi(\bs{r}) = \varphi_{xc}(\bs{r})
               +  \int d\bs{r}' u(\bs{r}-\bs{r}')n_{in}(\bs{r}')
\end{equation}
and
\begin{equation}\label{xc_poten}
\varphi_{xc}(\bs{r})=\frac{\delta E_{xc}[n_{in}] }{\delta n_{in}(\bs{r})} \;.
\end{equation}
Equations (\ref{varE}),(\ref{varN}) are precisely the same equations one
obtains starting with the Hamiltonian,
\begin{equation}\label{KS_hamiltonian}
 H = \sum_i^N
 \frac{\bs{p}^2_i}{2}-\frac{\bs{P}^2}{2N}+\sum_{i}\varphi(\bs{r}_i-\bs{R}) \;.
\end{equation}
Unfortunately this Hamiltonian cannot be separated into single particle
orbitals like the original KS Hamiltonian. However, since the center of mass
coordinate is a slow coordinate we can replace all the center of mass
operators by their expectation values.
In order to do so, let us assume that $\varphi$ is a smooth function that can be
expanded in the following manner,
\begin{equation}\label{expand_phi}
 \varphi(\bs{r}-\bs{R})\cong \varphi(\bs{r})
          - R_a \partial_a \varphi(\bs{r})
          +\frac{1}{2} R_a R_b \partial^2_{ab}\varphi(\bs{r})
          - \ldots \;,
\end{equation}
where 
\begin{eqnarray} 
  \partial_a \varphi(\bs{r})& = & \frac{\partial \varphi(\bs{r})}{\partial r_a} \cr
  \partial^2_{ab}\varphi(\bs{r}) & = & 
            \frac{\partial^2\varphi(\bs{r})}{\partial r_a \partial r_b} \;,
\end{eqnarray}
and summation over the spatial directions $a,b=\{x,y,z\}$ is assumed.
Replacing the center of mass terms by their expectation values 
and choosing a coordinate system such that $\bra \bs{R}\ket=0$, we obtain
\begin{eqnarray}\label{H_approx}
  H & \cong & \sum_i^N
 \frac{\bs{p}^2_i}{2}-\frac{\bra\bs{P}^2\ket}{2N} 
 -N\bra R_a^2 \ket \partial^2_{aa} \varphi|_0
 \cr
  & + & 
  \sum_{i}\left\{ \varphi(\bs{r}_i)
    +\frac{1}{2}\bra R_a R_b \ket
         \partial^2_{ab}\varphi(\bs{r}_i)
     \right\}
 \;. 
\end{eqnarray}
retaining only the leading center of mass corrections. The
$\partial^2_{aa}\varphi|_0$ 
correction accounts for a linear term in $\nabla \varphi$ which upon summation
over 
all particles is proportional to $\bs{R}$.
Clearly this Hamiltonian is a sum of N single--particle Hamiltonians
leading to the KS type orbitals in the Hartree approximation
\begin{equation}\label{orbital}
 \left[ -\frac{1}{2}\nabla^2 + \varphi(\bs{r})
    +\frac{1}{2}\bra R_a R_b \ket \partial^2_{ab}\varphi(\bs{r})\right]\psi_i
    = \epsilon_i \psi_i \;.
\end{equation}
To be more concrete let us consider the local density approximation (LDA) 
in which it is assumed that
$\varphi_{xc}(\bs{r})=\varphi_{xc}(n_{in}(\bs{r}))$. In this approximation the
leading correction to the single particle potential due to
the center of mass is
\begin{equation}
\delta \varphi(\bs{r}) = \frac{1}{2}\bra R_a R_b \ket 
                         \partial^2_{ab} \varphi(\bs{r}) 
\end{equation}
where
\begin{eqnarray} 
\partial^2_{ab} \varphi(\bs{r}) & = & 
    \frac{d \varphi_{xc}}{d n_{in}} \partial^2_{ab}n(\bs{r})
    + \frac{d^2 \varphi_{xc}}{d n_{in}^2}
       \partial_{a}n(\bs{r})\partial_{b}n(\bs{r}) 
   \cr 
   & + & \int d\bs{r}' u(\bs{r}-\bs{r}')\partial^2_{ab} n(\bs{r}') \;.
\end{eqnarray}
Here $n(\bs{r})$ is the laboratory one--body density given by
\begin{equation}\label{sp_den}
n(\bs{r})=\sum_i^N |\psi_i(\bs{r})|^2 \;. 
\end{equation}
Equations (\ref{1b_poten}),(\ref{xc_poten}), (\ref{orbital})-(\ref{sp_den})
together with
\begin{equation}\label{RR}
 \bra R_a R_b \ket = \frac{\bra r_a r_b \ket}{N}
                   = \frac{1}{N^2} \sum_i^N \bra \psi_i | r_a r_b | \psi_i \ket
\end{equation}
form a closed set of equations to be solved self-consistently, just as in the
case of a system bounded by an external potential.

The nature of
the center of mass expansion depends on the system at hand. 
This point is most easily
demonstrated in the Hartree picture where the expectation value of all the odd
moments of $R_a$ can be set to zero and the second even moment is given
by  
\begin{eqnarray}\label{RRRR}
\lefteqn{ \bra R_a R_b R_c R_d \ket =}\cr
    & & \hspace{-5mm}
     \frac{1}{N^2}\left(
     \bra r_a r_b \ket \bra r_c r_d \ket
   + \bra r_a r_c \ket \bra r_b r_d \ket
   + \bra r_a r_d \ket \bra r_b r_c \ket
   \right)
    \cr & & \hspace{-5mm}
     + O(\frac{1}{N^3})
 \;.
\end{eqnarray}
From Eqs. (\ref{RR}) and (\ref{RRRR}) it is clear that the leading orders in 
the expansion (\ref{expand_phi}) can be regarded as a Taylor series in 
$\bra r^2 \ket/N$, and that in the limit
$N \longrightarrow \infty$ the original KS equations are recovered.
For a collapsing system where the size of the system does not depend
on the number of particles the expansion
(\ref{expand_phi}) behaves as a $1/N$ series. For non-collapsing systems, such
as nuclei, the rms radius grows as $\sqrt[3]{N}$ and the expansion
converges much slower as $1/\sqrt[3]{N}$, unless the $r$-dependence of the
potential scales as the rms matter radius, i.e.
 $\varphi \approx \varphi(\bs{r}/\sqrt[3]{N}a)$. 

{\it Example --} 
As an example for our procedure consider a system of N fermions moving in a
harmonic oscillator potential,
\begin{equation}\label{HO_hamiltonian}
 H = \sum_i^N \frac{\bs{p}_i^2}{2}+\sum_{i}\frac{1}{2}\omega^2 \bs{r}_i^2 \;.
\end{equation}
Using the relation
\begin{equation}
 \sum_{i}^N (\bs{r}_i-\bs{R})^2=\sum_i^N \bs{r}_i^2-N \bs{R}^2
\end{equation}
we can rewrite this Hamiltonian as a sum of a center of mass term
\begin{equation} \label{HO_cm}
  H_{cm} = \frac{\bs{P}^2}{2N}+\frac{1}{2}N\omega^2 \bs{R}^2 \;,
\end{equation}
and an internal Hamiltonian written in the form (\ref{KS_hamiltonian}),
\begin{equation}
  H_{in} =  \sum_i^N \frac{\bs{p}^2_i}{2}-\frac{\bs{P}^2}{2N}
 + \frac{1}{2}\sum_{i}^N \omega^2 (\bs{r}_i-\bs{R})^2 \;.
\end{equation}
Following Eq. (\ref{H_approx}), we evaluate the center of mass corrections
\begin{eqnarray}
  N\bra {R}_a R_a\ket \partial^2_{aa} \varphi|_0 & = &
  N \bra \bs{R}^2\ket \omega^2  \;,
  \cr
  \frac{1}{2}\bra R_a R_b \ket \partial^2_{ab}\varphi(\bs{r}_i) & = &
  \frac{1}{2} \bra \bs{R}^2\ket \omega^2 \;,
\end{eqnarray}
and get the approximated internal Hamiltonian
\begin{equation}\label{HO_approx}
 H_{in} \cong \sum_i^N \frac{\bs{p}_i^2}{2}
        - \frac{\bra \bs{P}^2 \ket}{2N}-\frac{1}{2}N\omega^2 \bra\bs{R}^2 \ket
        +\sum_{i}\frac{1}{2}\omega^2 \bs{r}_i^2 \;.
\end{equation}
Comparing (\ref{HO_approx}) with (\ref{HO_cm}) it is clear that for the
harmonic oscillator case the 
corrections to the internal Hamiltonian ensure the right cancellation of the
center of mass energy.  

{\it Discussion --} 
It is already known for some time that the HK theorem can be generalized to
any Hermitian operator, including the intrinsic density. The challenge however
is to find a useful KS like procedure that reduces the self-bound
many-body system into a set of single quasi-particle orbitals. From the onset
it is clear that such orbitals break the translational symmetry, and therefore
some approximations are called for. In this note we have
demonstrated that starting with the HK theorem for the intrinsic density one
can reach this goal by treating the center of mass coordinate as an adiabatic
variable. 
This approximation recovers the KS equation in the limit $N\longrightarrow
\infty$ and includes center of mass recoil effects in the orbitals equations
and to the ground state energy. We have demonstrated that for the case of
$N$-particles moving in an harmonic-oscillator potential this procedure yields
the right center of mass correction.

We argue that the importance of the center of mass corrections might vary
for different systems. Regardless whether the effect is small or large the
current 
approach helps bridge the gap between the DFT and its application to finite 
self--bound systems.

I wish to thank W. Kohn, G. F. Bertsch, and B. R. Barrett for 
useful discussions and help 
during the preparation of this work.
This work was supported by the Department of Energy Grant
No. DE-FG02-00ER41132. 



\begin{thebibliography}{99}
\bibitem{Engel06} J. Engel, Phys. Rev. C {\bf 75}, 028501 (2007).
\bibitem{Giruad06} B. G. Giruad, B. K. Jennings, and B. R. Barrett,
           arXiv:0707.3099 (2007).
\bibitem{Valiev97} M. Valiev and G. W. Fernando, arXiv: cond-math/9702247
  (1997). 
\bibitem{HK} P. Hohenberg and W. Kohn, Phys. Rev. {\bf 136}, B864 (1964).
\bibitem{KS} W. Kohn and L. J. Sham, Phys. Rev. {\bf 140}, A1133 (1965).

\end{thebibliography}
\end{document}